\documentclass{emulateapj}
\usepackage{graphicx}

\newcommand{\beq}{\begin{eqnarray}}
\newcommand{\eeq}{\end{eqnarray}}
\newcommand{\bfig}{\begin{figure}}
\newcommand{\efig}{\end{figure}}
\def \lsi {LS I $+61^{\circ}303$}
\def \gam {$\gamma$}

\usepackage{color}
\usepackage{amsmath}
\usepackage{ulem}

\shorttitle{Multi-wavelength Characteristics of \lsi}
\shortauthors{Saha et al.}
\begin{document}

\title{The Multi-wavelength Characteristics of the TeV Binary \\ \lsi }
\author{L. Saha\altaffilmark{1}, V. R. Chitnis, A. Shukla\altaffilmark{2}, A. R. Rao and B. S. Acharya}
\affil{Tata Institute of Fundamental Research,
Homi Bhabha Road, Colaba, Mumbai 400 005, India}
\email{labsaha@ncac.torun.pl, vchitnis@tifr.res.in}
\altaffiltext{1}{Present address :  Nicolaus Copernicus Astronomical Center, Torun, Poland}
\altaffiltext{2}{Present address : ETH Zurich, Institute for Particle Physics, Otto-Stern-Weg 5, 8093 Zurich, Switzerland}

\begin{abstract}
We study the  characteristics of the TeV binary \lsi\ in radio, soft X-ray, hard X-ray, and gamma-ray 
(GeV and TeV) energies. The long term variability characteristics are examined as a function of the 
phase of  the binary period of 26.496 days as well as  the phase of the super orbital period of  1626
days, dividing the observations into a matrix of 10$\times$10 phases of these two periods.  It is 
found that the long term variability can be described by a sine function of the super orbital period, 
with the phase and amplitude systematically varying with the binary period phase. We also find a definite 
wavelength dependent change in this variability description. To understand the radiation mechanism, 
we define three states in the orbital/ super orbital phase matrix and examine the wide band spectral 
energy distribution.  The derived source parameters indicate that the emission geometry is dominated 
by a jet structure showing a systematic variation with the orbital/ super orbital period. We suggest 
that \lsi\ is likely to be a micro-quasar with a steady jet.

\end{abstract}

\maketitle

\section{Introduction}

\lsi, a Galactic high-mass X-ray binary system located at a distance of 2 
kpc \citep{Frail-1991}, is detected in the energy range from radio to 
\gam-rays exhibiting strong variable emission. It consists of a B0 
main-sequence star with a circumstellar disk (i.e a Be star) and a compact 
object of unknown nature. The orbital period of the system is estimated
to be $P_{orb}$ = 26.496 days and it also exhibits a long term periodic 
variation with a superorbital period of  $P_{sup}$ = 1667 
days \citep{Gregory-2002,Massi-2013,Massi-2015}. However, very recently the 
superorbital period is estimated to be 1626 days using 37 years of radio data 
\citep{Massi_2016}.
The zero orbital phase 
corresponds to $T_{0,orb} =  2443366.775 + nP_{orb}~JD$. According to the 
most recent radial velocity measurements, the orbit is elliptical with
eccentricity of e = 0.537 $\pm$ 0.034 and periastron passage occurring around phase 
$\phi$ = 0.275, apastron passage at $\phi$ = 0.775, superior conjunction 
at $\phi$ = 0.081, and inferior conjunction at $\phi$ = 0.313 
\citep{Aragona-2009}.

High angular resolution VLBI radio data has shown the presence of 
high-energy particle outflow possibly related to jet-like ejection on 
the time scale of a orbital period \citep{Paredes-1998, Massi-2004}.  
However, the observed morphological changes in the data
collected at different epochs reported by \cite{Dhawan-2006} 
support a scenario of binary pulsar. Recent
detailed VLBA radio images, obtained by reprocessing same
data-set,  through the orbital period established the
presence of one sided and double sided radio structures
supporting a precessing microquasar model \citep{Massi-2012}.

Long-term monitoring of the source during 2007-2011 by Proportional 
Counter Array (PCA) onboard Rossi X-ray Timing Explorer (RXTE) 
established the superorbital modulation in X-rays and a shift of 
superorbital phase by 0.2 between radio and X-ray data \citep{Li-2012, 
Chernyakova-2012}.  Very recently, superorbital modulation at MeV--GeV 
\gam-rays in the apastron phase (0.5 --1.0) has been established by 
Fermi-LAT \citep{Ackermann-2013} based on the data taken during 2008 
August 4 to 2013 March 24.  

This source has often shown complex behaviour in very high energy \gam-rays.
\lsi\ was first observed at TeV energies by the MAGIC telescope system 
during  2005 October -- 2006 March 
with a significance of 8.7$\sigma$ in the orbital phase 0.4--0.7, 
establishing it as a \gam-ray binary
\citep{Albert-2006}. The VERITAS observations carried out during 2006 
September -- 2007 February confirmed TeV emission from this
source \citep{Acciari-2008}.  However, further observations of the source 
by both MAGIC and VERITAS have shown different flux levels 
\citep{Acciari-2011,Aleksic-2012,Aliu-2013}. These observations at TeV 
energies show that the source behaves differently in different orbital 
cycles suggesting a variable nature of the source. Variability of the 
source almost in all wavelengths could possibly be related to superorbital 
modulation of the fluxes, which has been shown at radio, X-ray and 
MeV--GeV \gam-rays detected by Fermi-LAT  \citep{Gregory-2002,Li-2012, 
Chernyakova-2012,Ackermann-2013}. Hence long term multiwaveband study of
this source can provide an important observational support for unveiling
the nature of the source and the emission mechanisms.

With this motivation, we have studied the radio, X-ray and \gam-ray data from
this source collected over a period longer than the superorbital period. We
have studied the variation of flux as a function of orbital and superorbital
phases. We have also studied multiwaveband Spectral Energy Distribution
(SED) of the source in some of the phases. This paper is organized as follows:
In section 2, the data set used for these studies and analysis procedure is
described. Variation of the flux with the orbital and the superorbital phases is
discussed in section 3. The SEDs and their interpretation in terms of
microquasar model are given in section 4 followed by a discussion and 
conclusions in section 5.

\section{Multiwaveband Data and Analysis}

In the last few years, \lsi\ has been observed extensively by various instruments.
In the present work, we have used data from radio, X-ray and \gam-ray bands.
Radio data used here is from \cite{Richards-2011} and \cite{Massi-2015}. These are
15 GHz observations from 40 m single-dish telescope at Owens Valley Radio 
Observatory (OVRO). Data on \lsi\ were collected during MJD 54908.8 -- 56795.0
(2009 March -- 2014 May). Observations were carried out approximately 
twice a week. 

X-ray data were obtained from PCA onboard RXTE and X-ray Telescope or XRT
onboard Swift. The PCA is an array of five identical 
Xenon filled proportional counter units (PCUs) \citep{Bradt_1993} covering an energy 
range from 2 to 60 keV with a total collecting area of 6500 cm$^2$. 
Data were collected over the period MJD 50143 -- 55924 and
standard analysis procedure was used to generate PCA
light curves over the energy range of 2 --­ 9 keV.

The XRT onboard Swift consists of a grazing incidence Wolter I
telescope which focuses X-rays on a CCD \citep{Burrows_2005}. This
instrument has an effective area of 110 cm$^2$, 23.6 arcmin field
of view (FOV) and 15 arcsec resolution (half-power diameter).
It covers an energy range of 0.2 to 10 keV. 
Swift-XRT light curves were obtained from the site 
\footnote{$http://www.swift.ac.uk/user\_objects/$}.
Data is collected over the period of MJD 53980 -- 57039 (2006 September 2 
-- 2015 January 17). Details of the procedure used for generating
these light curves is given in \cite{Evans_2007}.

High energy \gam-ray data  are obtained from Large Area Telescope (LAT)
onboard Fermi. The Fermi-LAT is a pair production telescope \citep{Atwood_2009} 
covering energy range of 20 MeV to 300 GeV with a FOV of 
$\ge$ 2.5 sr. The data taken over the period MJD 54682.9 -- 57145.9
(2008 August 4 -- 2015 May 3) were analysed in the present work. Circular
region of interest (ROI) with radius 15$^\circ$ centred at the position 
of RA(J2000) = 02$^h$ 40$^m$ 34$^s$ and Dec(J2000) = 61$^\circ$ 15$'$ 25$''$
was used for extracting the data. Fermi Science Tools (FST-v10r0p5) with event 
class Pass 8 data 
were used for Galactic point source analysis. Since the Earth's limb is 
a strong source of background $\gamma$-rays, they were filtered out with a 
zenith-angle cut of 100$^\circ$.  A python based software tool {\it enrico} 
\citep{enrico-2013} was used to do standard binned likelihood analysis. 
The \gam-ray events in the data were binned in 8 logarithmic bins in the energies
between 300 MeV and 300 GeV. Since the point-spread function (PSF) of 
LAT is large, the sources from outside of the ROI may  contribute at low energies affecting 
true estimates of the fluxes for the sources considered in this analysis. In order to account  
for this, exposure map was expanded  by another 10$^\circ$ outside the ROI, for all events, 
as suggested by \citet{Abdo-2009}.

We studied the spectral properties of  the \gam-ray emission by comparing 
the observational results with the models of the sources present in the ROI. To get the 
best-fit model parameters, the spatial distribution and spectral  models of the sources 
are  convolved with the instrument response function (IRF) and exposure of the observation. 
In this work, we used newly introduced IRF version \textit{ P8R2\_SOURCE\_V6}. There are 85 
point-like sources  and some diffuse background sources from the 3rd Fermi-LAT catalog 
located in the ROI. In order to account for the emission from  background sources, we 
considered two component background model: diffuse Galactic emission (gll\_iem\_v06.fits) 
and isotropic emission component (iso\_P8R2\_SOURCE\_V6\_v06.txt) consisting of emission 
from extra galactic background, unresolved sources and instrumental background.

The binned likelihood analysis was used for both background and source 
modelling using {\it gtlike} tool of FST. Spectral parameters for the source outside 
the 3$^\circ$ region centred at the \lsi\ position were kept fixed. 
However, parameters except normalization  for the point-like background 
sources were fixed or  varied based on their strength and distance from the center of the 
ROI. Light curve was generated over the energy range of 300 MeV -- 300 GeV.

For Very High Energy (VHE) or TeV band, published data collected during 2005-2011
from MAGIC \citep{Albert-2009} and VERITAS \citep{Acciari-2008,Acciari-2011,Aliu-2013} 
experiments are used. These are the ground
based atmospheric Cherenkov experiments located in La Plama and Arizona, respectively.

\section{Multiwaveband Flux Variation}

\begin{figure*}[t]
\begin{tabular}{cc}
\centering
\includegraphics[scale=0.32,angle=-90]{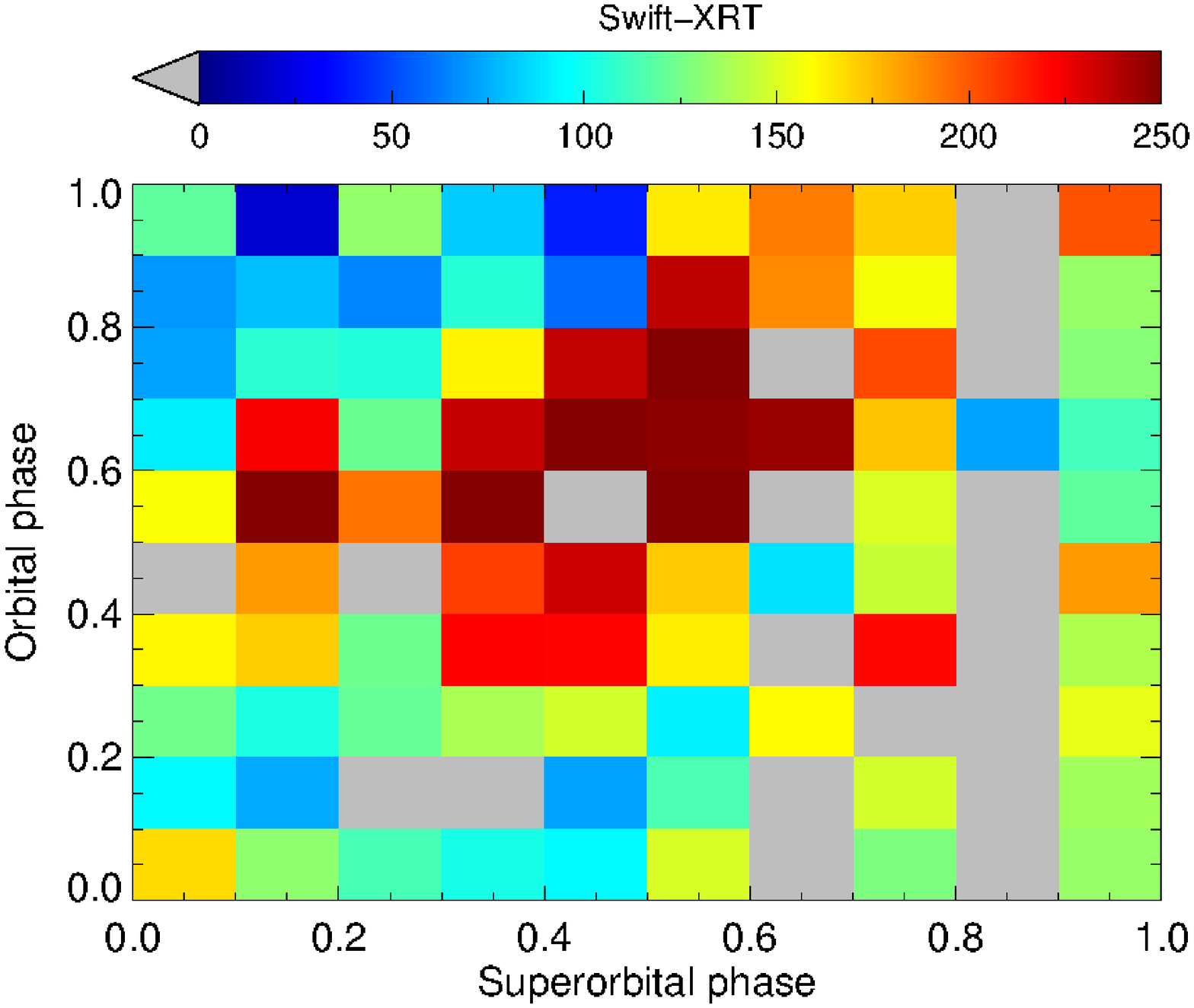}
\includegraphics[scale=0.32,angle=-90]{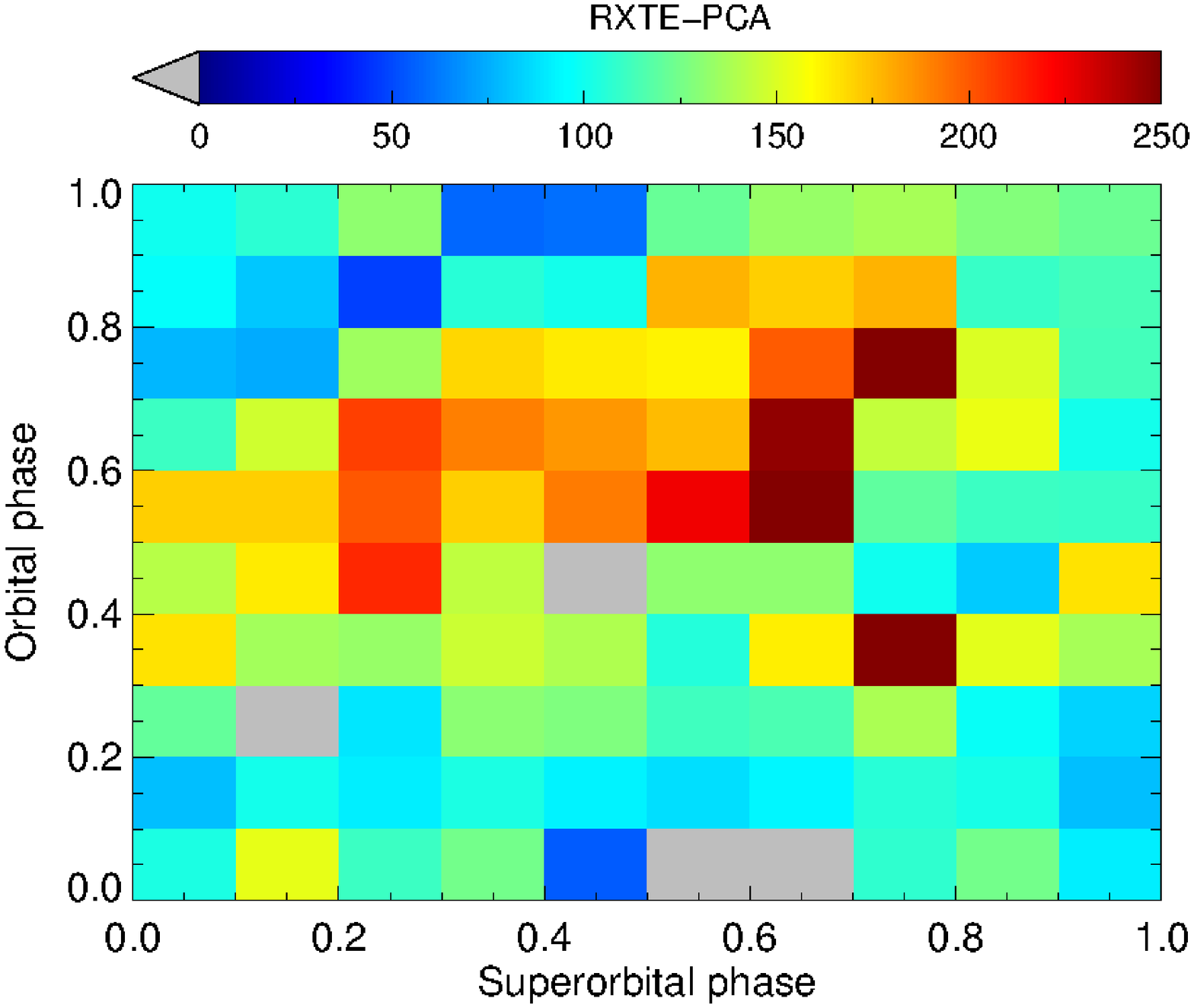}
\end{tabular}
\\
\begin{tabular}{cc}
\centering
\includegraphics[scale=0.32,angle=-90]{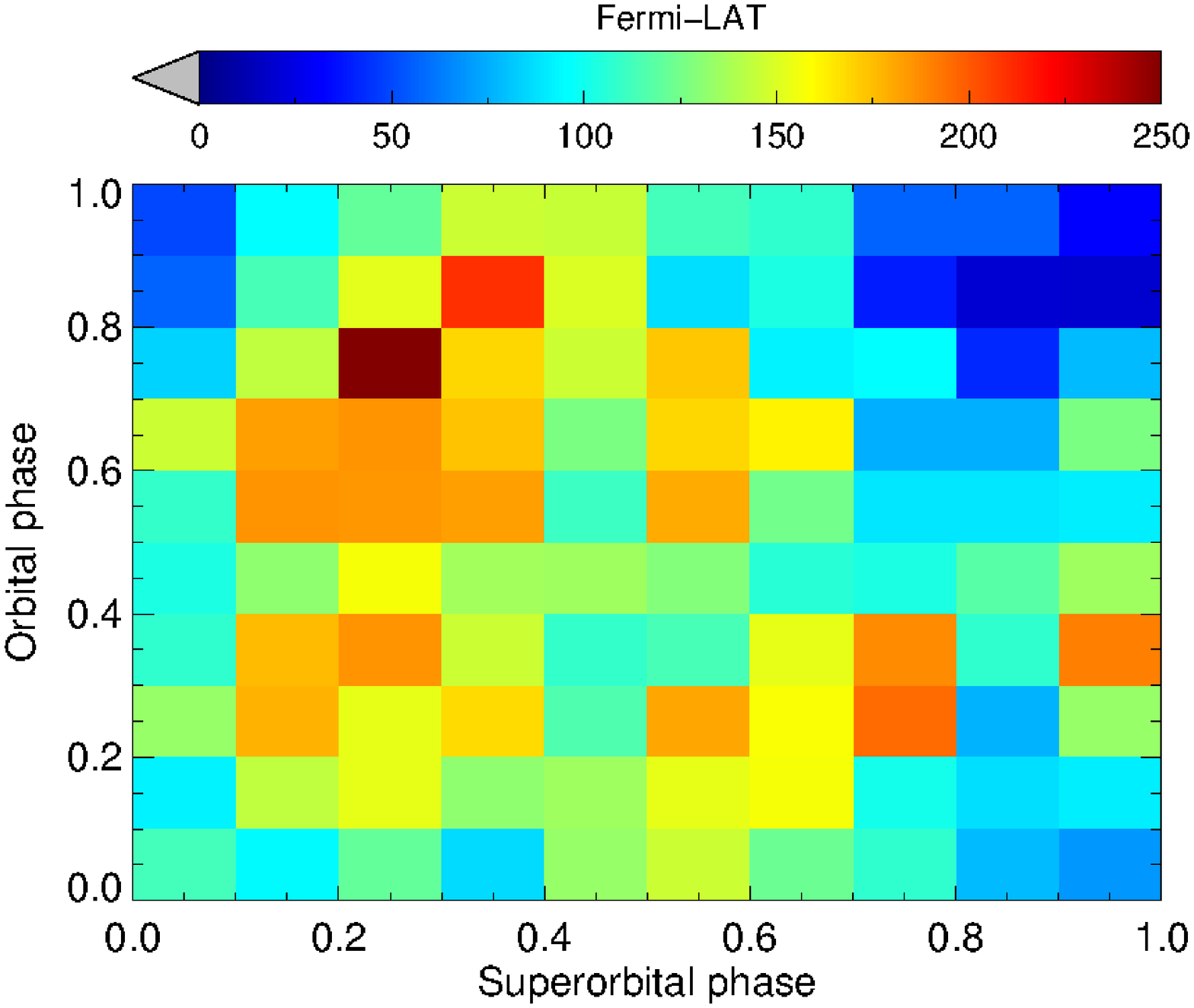}
\includegraphics[scale=0.32,angle=-90]{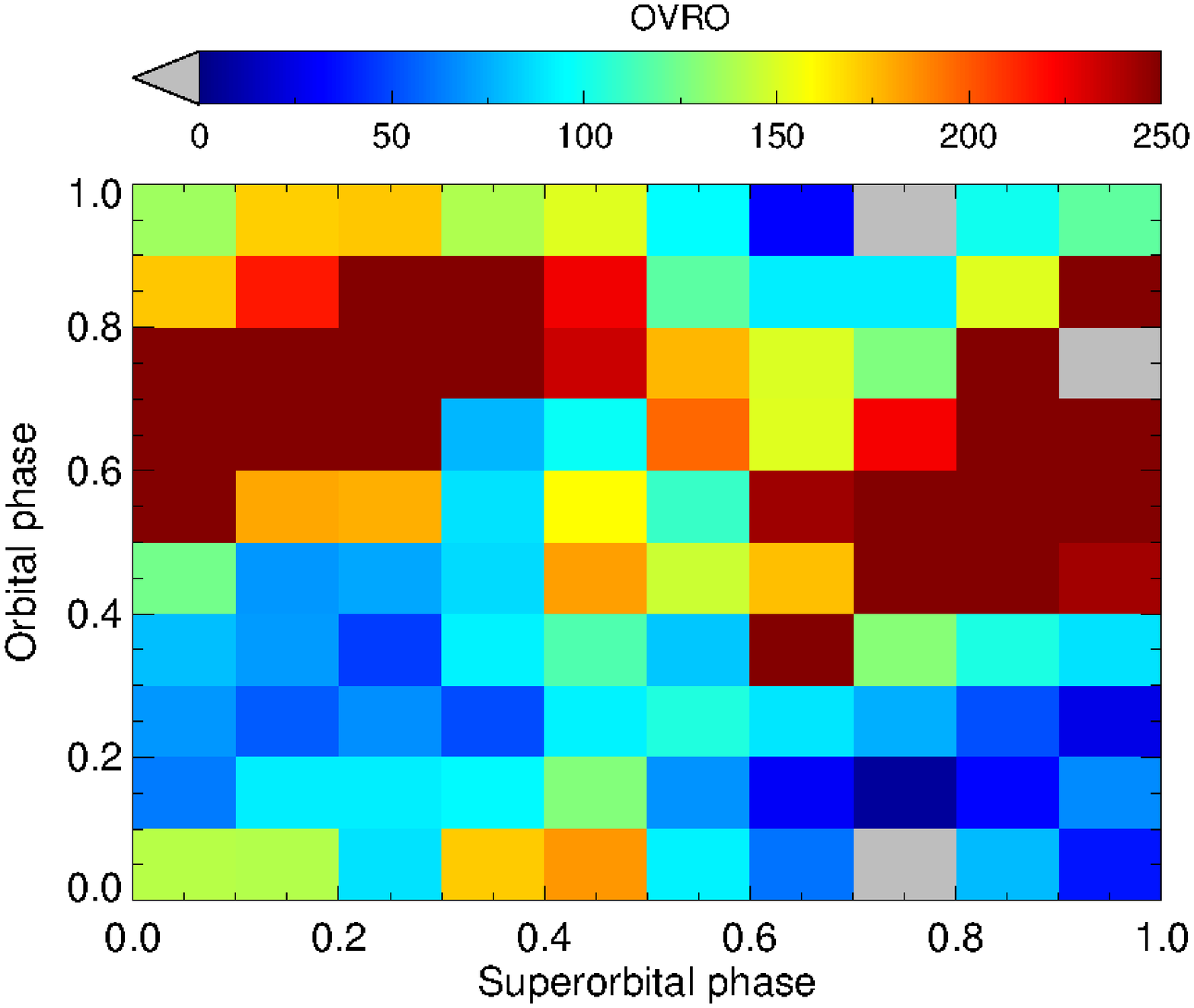}
\end{tabular}
\\
\begin{tabular}{cc}
\centering
\includegraphics[scale=0.32,angle=-90]{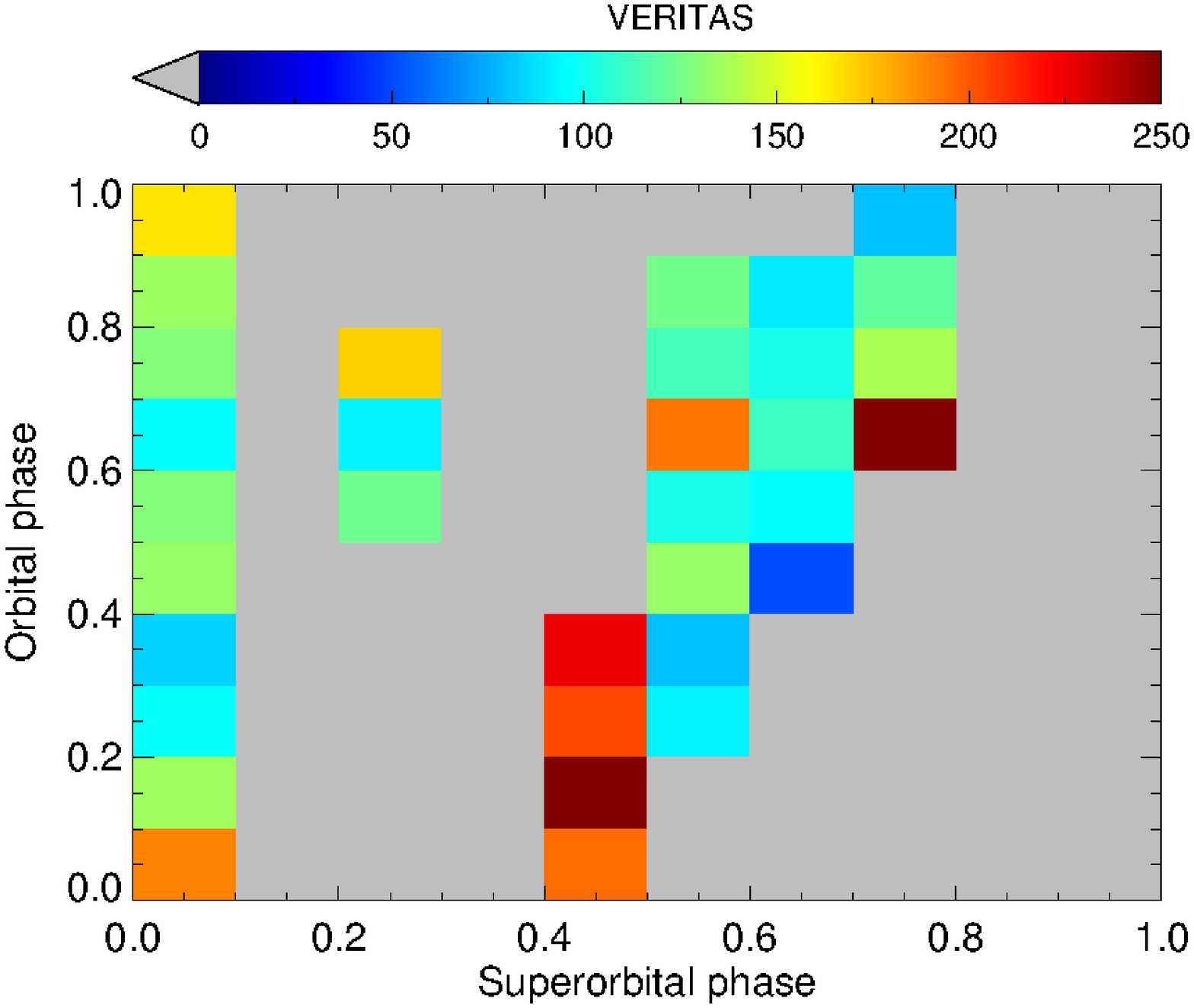}
\includegraphics[scale=0.32,angle=-90]{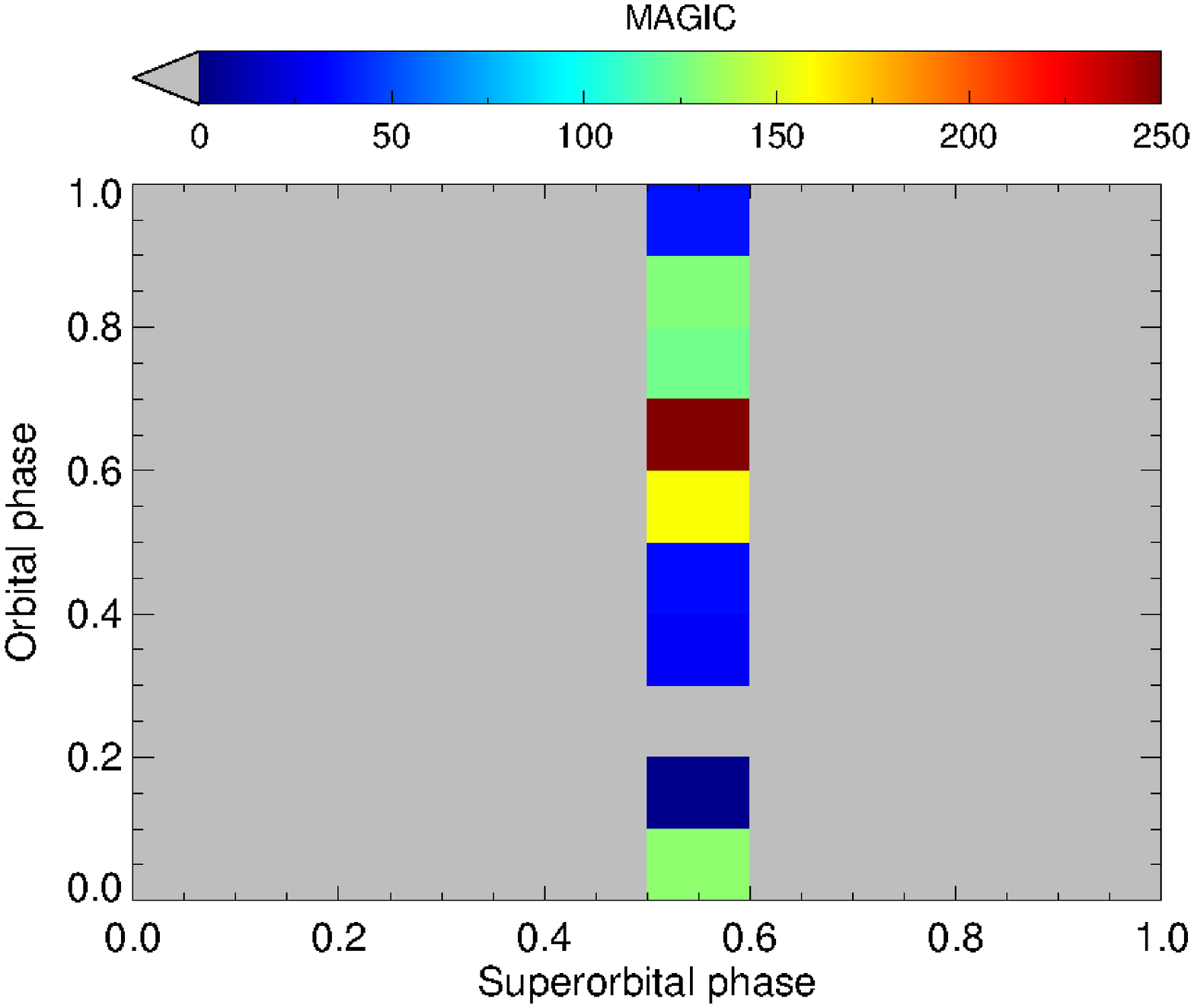}
\end{tabular}
\caption{Multiwaveband flux as a function of orbital and superorbital
phases. Top panels show X-ray flux from XRT data in the left panel and PCA
data in the right panel. Middle left panel corresponds to $\gamma$-ray data 
from Fermi-LAT and middle right panel shows radio data from OVRO. Bottom
panels correspond to VHE $\gamma$-ray flux from VERITAS (left panel)
and MAGIC (right panel). Flux values in each panel are normalized setting
median flux to 125, i.e. the middle of the scale.}
\label{fig:density-plot}
\end{figure*}

In order to study the variation of flux as a function of orbital and superorbital
phases, datasets from various wavebands were folded in 10 superorbital phase
bins using the ephemeris given by \cite{Massi_2016}. Each of these bins
correspond to 163 days. Further, in each phase bin, 10 orbital phase bins were 
generated. Average flux was estimated in each of these 10 $\times$ 10 phase bins. 
X-ray fluxes from Swift-XRT and RXTE-PCA in 10 $\times$ 10 orbital vs superorbital 
bins are shown in the top left and the top right panels of Fig. \ref{fig:density-plot}, respectively.  
These panels 
show a definite pattern in the variation of flux over the orbital and the superorbital 
phases for both XRT and PCA data. It can be seen from the figure that the source 
is bright in the orbital phase range $\sim$ 0.4 -- 0.8 while the corresponding superorbital 
phase is at $\sim$ 0.3 -- 0.8. The highest flux in each of the orbital cycles 
shifts towards apastron passage as the superorbital phase value increases. Similar
plots generated for the \gam-ray data from Fermi-LAT and 15 GHz radio data from
OVRO are given in the middle panels of Fig. \ref{fig:density-plot}. 
As noted by earlier 
studies, there is a definite shift in the pattern for the radio data: the maximum flux
is at the same orbital phase range (0.4 -- 0.8) as in the X-ray data, but the super 
orbital phase range is shifted to 0.7 to 1.4.
The Fermi-LAT data shows some indication of enhanced emission in the orbital phase 0.4 to 0.8, but
unlike the other wavebands, the enhancements in the super orbital phases are not very clear.
Plots for VHE \gam-ray data from VERITAS and MAGIC
are shown in bottom panels of Fig. \ref{fig:density-plot}. But the data-set is not 
extensive enough to detect any trend in this case.

\begin{figure*}[t]
\centering
\begin{tabular}{cc}
\includegraphics[scale=0.45]{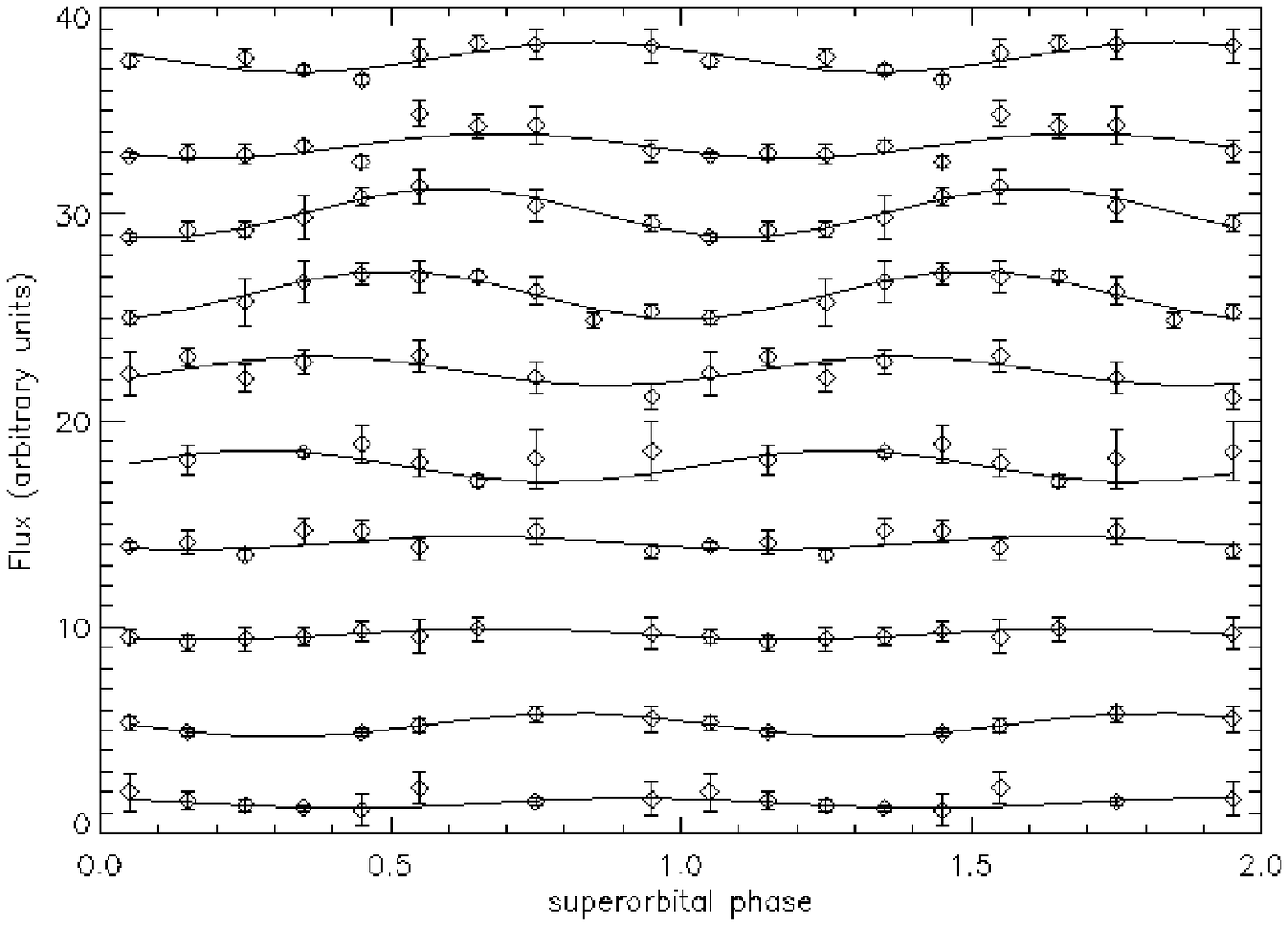}
\includegraphics[scale=0.45]{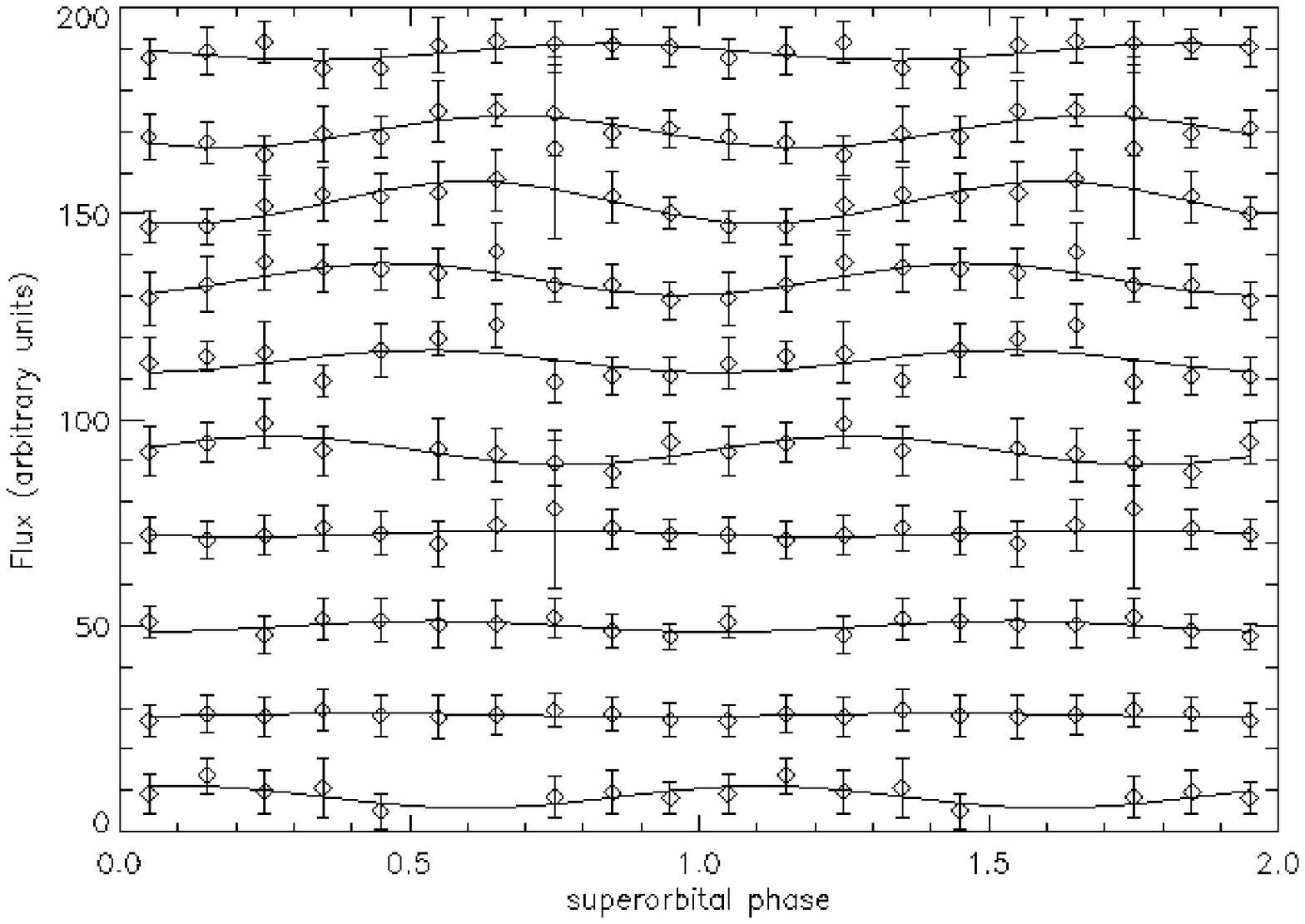}
\end{tabular}
\\
\begin{tabular}{cc}
\includegraphics[scale=0.45]{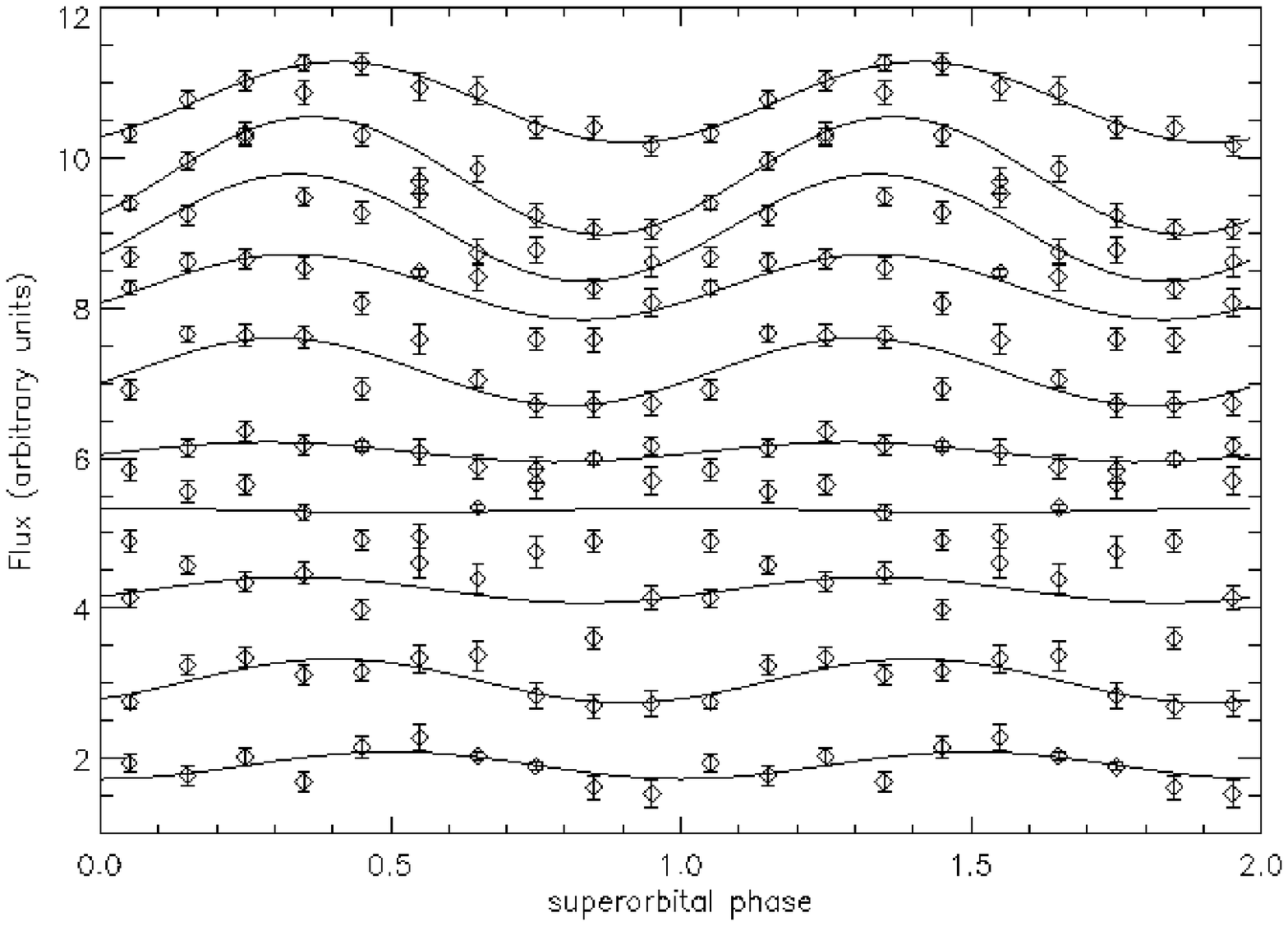}
\includegraphics[scale=0.45]{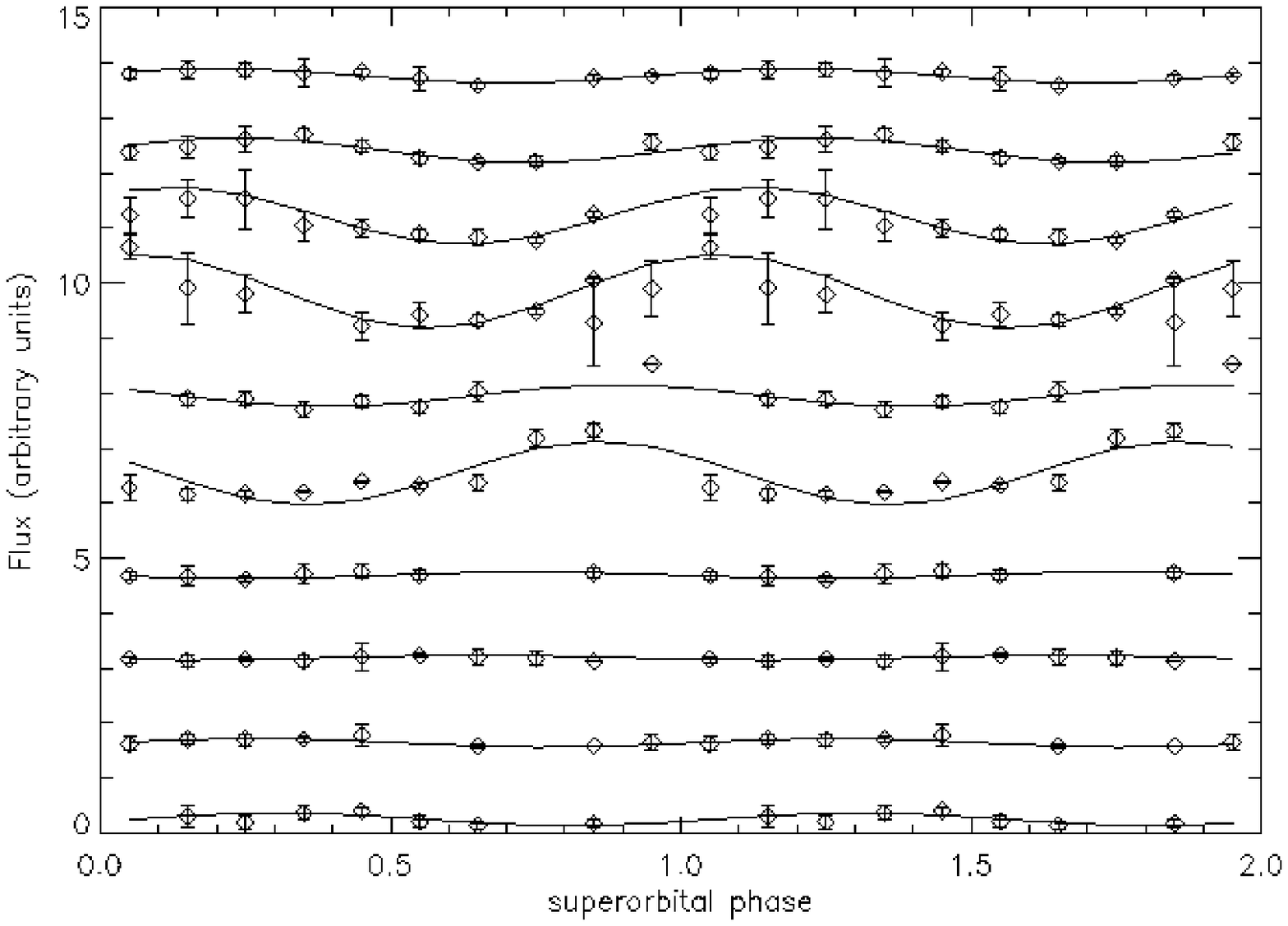}
\end{tabular}
\caption{Variation of flux with superorbital phase for XRT (top left),
PCA (top right), Fermi-LAT (bottom left) and OVRO (bottom right). In
each panel curves from bottom to top correspond to orbital phases 
0-0.1, 0.1-0.2, .. , 0.9-1.0. These curves are shifted along Y-axis 
for the sake of clarity.}
\label{fig:xrt-lc2}
\end{figure*}

To investigate this aspect further, variation of the flux as a function of superorbital
phase was studied in various orbital phase bins. Variation of the X-ray count rates from
Swift-XRT with superorbital phase is shown in the top left panel of Fig. \ref{fig:xrt-lc2}. 
Similar plots for RXTE-PCA, Fermi-LAT and OVRO are shown in the top right, bottom
left and bottom right panels of the same figure, respectively. 
In each panel, different curves from bottom to top 
correspond to orbital phases 0-0.1, 0.1-0.2, .. , 0.9-1.0. Curves are shifted with 
respect to each other for the sake of clarity. Error bars correspond to the standard 
deviation in each bin. To parameterize this variation, data are fitted with a constant 
and alternatively with a sine function of the form 
$f(t) = f_o + A \times sin(\phi_s - \phi_o)$, where $f_o$, $A$ and $\phi_o$ are model
parameters and $\phi_s$ is the superorbital phase. Sine function with a period of 1626 
days gives a better fit than the constant. 
It can be seen from the figure that there is a definite 
shift in the superorbital phase for the peak flux with respect to the  orbital phase.
Phase at the peak of the function, peak function value and the ratio of the maximum to the minimum 
function values are listed in Table ~\ref{tab:super_peak_orb}. Results are given only for 
the cases where modulation is seen clearly in Fig. ~\ref{fig:xrt-lc2}. 
Fig. ~\ref{fig:super_peak_orb} shows 
these results graphically, where superorbital phase for peak of the function value is 
plotted as a function of orbital phase bins for XRT, PCA, Fermi-LAT and OVRO data.
This figure clearly shows the trend of increasing superorbital phase for peak as a
function of orbital phase near apastron. The wavelength dependent phase difference 
between superorbital phase for given orbital phase bin is also evident. This
difference remains more or less constant in various orbital phase bins near apastron.

%\clearpage
\begin{deluxetable*}{ccccccccccccc}
\tabletypesize{\scriptsize}
%\rotate
\tablecaption{Peak flux and corresponding superorbital (SO) phase from sine function fit in various orbital phase bins \label{tab:super_peak_orb}}
\tablewidth{0pt}
\tablehead{
\colhead{Orbital} & \multicolumn{3}{c}{XRT} & \multicolumn{3}{c}{PCA} &\multicolumn{3}{c}{Fermi-LAT} &\multicolumn{3}{c}{OVRO} \\
\colhead{Phase} & \colhead{SO} & \colhead{peak flux} & \colhead{ratio} & \colhead{SO} & \colhead{peak flux} & \colhead{ratio} & \colhead{SO} & \colhead{peak flux} & \colhead{ratio} & \colhead{SO} & \colhead{peak flux} & \colhead{ratio} \\
\colhead{} & \colhead{phase} & \colhead{($10^{-1} ph$} & \colhead{max} & \colhead{phase} & \colhead{($10^{-1} ph$} & \colhead{max} & \colhead{phase} & \colhead{($10^{-1} ph$} & \colhead{max} & \colhead{phase} & \colhead{($10^{-1} ph$} & \colhead{max} \\
\colhead{} & \colhead{at peak} & \colhead{$ cm^{-2}~s^{-1}$)} & \colhead{/min)} & \colhead{at peak} & \colhead{$ cm^{-2}~s^{-1}$)} & \colhead{/min)} & \colhead{at peak} & \colhead{$ cm^{-2}~s^{-1}$)} & \colhead{/min)} & \colhead{at peak} & \colhead{$ cm^{-2}~s^{-1}$)} & \colhead{/min)}\\
}
\startdata
0.0--0.1     &   -    &   -   & -     & -    &   -    &  -    & -   &    -   &  -     &0.30  &  4.19 &23.88\\
0.1--0.2     &   -    &   -   & -     & -    &   -    &  -    & -   &    -   &  -     &0.26  &  2.22 &3.25\\
0.2--0.3     &   -    &   -   & -     & -    &   -    &  -    & -   &    -   &  -     &0.58 &   2.32 &1.86 \\
0.3--0.4     &   -    &   -   & -     & -    &   -    &  -    & -   &    -   &  -     &0.70 &   2.51 &2.01\\
0.4--0.5     &  0.28  &  2.79 &3.16   &0.26  &  1.60  & 1.80  &0.28 &   2.21 & 1.13   &0.80  &  8.73 &7.75\\
0.5--0.6     &  0.38  &  3.12 &1.86   &0.54  &  1.83  & 1.68  &0.30 &   2.60 & 1.53   &0.90  &  6.50 &2.58\\
0.6--0.7     &  0.50  &  3.26 &3.55   &0.46  &  1.75  & 1.70  &0.34 &   2.71 & 1.47   &0.06  &  9.31 &3.10\\
0.7--0.8     &  0.60  &  3.27 &3.79   &0.62  &  1.64  & 2.12  &0.34 &   2.78 & 2.06   &0.10  & 9.93 &2.85\\
0.8--0.9     &  0.60  &  2.53 &3.19   &0.66  &  1.46  & 2.50  &0.36 &   2.54 & 2.60   &0.30  &  6.41 &6.91\\
0.9--1.0     &  0.84  &  2.34 &2.71   &0.78  &  1.16  & 1.51  &0.42 &   2.28 & 1.89   &0.26 &   3.82 &1.82\\
\enddata
\end{deluxetable*}

\bfig[t]
\centering
\includegraphics[scale=0.35,angle=0]{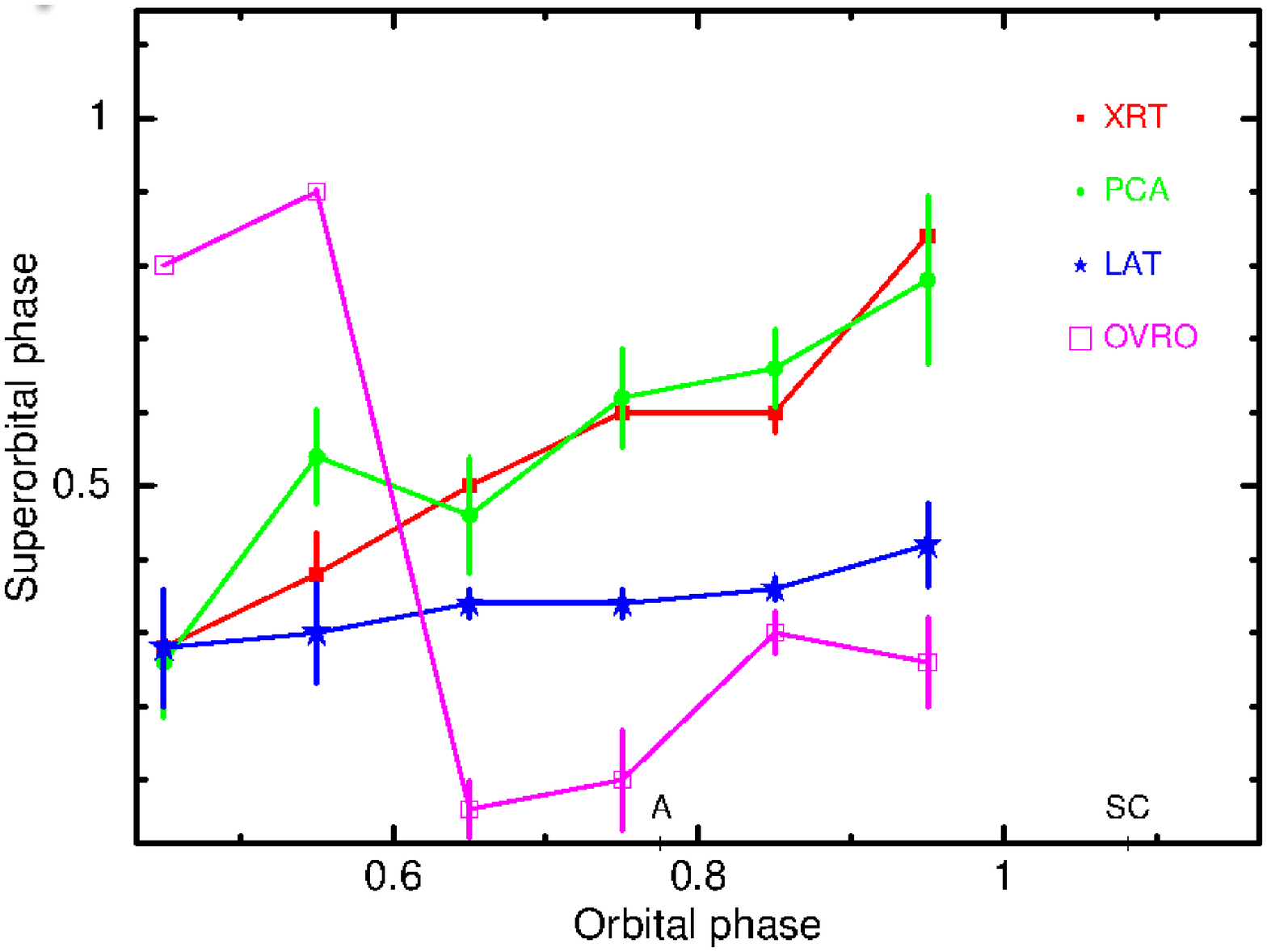}
\caption{Superorbital phase at peak flux from fitted sinusoidal function given in 
Table ~\ref{tab:super_peak_orb}
as a function of orbital phase bins for XRT, PCA, Fermi-LAT and OVRO data. Positions
for apastron (A) and superior conjunction (SC) are marked.}
\label{fig:super_peak_orb}
\efig

\section{Spectral Energy Distribution}\label{sec:sed}

We have investigated the spectral properties of the source at
different orbital and superorbital phases. Following Figure \ref{fig:density-plot}
 three different regions were chosen.
Two of the regions are bright in most of the wavebands and the third one is of low brightness. 
These regions are i) Super orbital phase: 0.3 -- 0.5 Orbital phase: 0.6 -- 0.8, 
ii) Super orbital phase: 0.5 -- 0.7 Orbital phase: 0.6 -- 0.8 and iii) Super orbital 
phase: 0.0 -- 0.2 Orbital phase, 0.0 -- 0.2 (hereafter state1, state2, and state3, 
respectively). X-ray and Fermi-LAT spectral data were analysed for these three regions.
Fermi-LAT analysis procedure is already described in section 2. We have analysed
spectral data from Swift-XRT and RXTE-PCA corresponding to the states mentioned above.
Some details of these observations are given in Table \ref{tab:xrt-pca-log}.
Dates for XRT and PCA observations for each of the three states are listed in the
table along with the total observation duration. 

In case of XRT we have fitted spectrum over the energy range of 0.3 to 10 keV.
Source and background photons were selected using the tool XSELECT. Data were recorded in
Photon Counting (PC) mode for these observations. Source photons were selected from a circular
region with the radius of 20 pixels (i.e. 47 arc-seconds), whereas 
nearby circular region with radius of 40 pixels was used for extracting 
background photons.
Events with grades 0-12 were selected in this analysis. 
The spectral data were rebinned using tool GRPPHA with 20 photons per bin. 
Standard response matrices and ancillary response files
were used.

In case of PCA, standard 2 data with time resolution of 16 s and  128 channels
of energy information were used. Data were analysed using HEASOFT (version 6.15). 
For each observation, data were filtered using the standard procedure
given in the RXTE Cook Book. The tool 'pcabackest' was used for generation
of background model, calibration files for 'faint' source (less than 40 ct/sec/PCU)
from RXTE GOF were used. To improve statistics, only
data from top layer of PCU2 was used.

%\clearpage
\begin{deluxetable*}{ccccc}
\tabletypesize{\scriptsize}
%\rotate
\tablecaption{Observation log for XRT and PCA \label{tab:xrt-pca-log}}
\tablewidth{0pt}
\tablehead{
\colhead{State} & \colhead{Instrument} & \colhead{Observation dates} & \colhead{Number of} & \colhead{Total duration}\\
\colhead{} & \colhead{} & \colhead{} & \colhead{observations} & \colhead{seconds}\\
}
\startdata
1     &  XRT       & 2010-10-22, 2010-11-18, 2010-12-17, 2014-10-18, & 13 &15805 \\
Superorb. phase : 0.3--0.5       &            & 2014-10-20 - 2014-10-23, 2014-11-14, 2014-11-15,   & &     \\
Orbital phase :  0.6--0.8      &            &  2014-12-11 - 2014-12-13   &&  \\
      &   PCA      & 2010-02-25, 2010-02-28, 2010-03-24, 2010-03-28,  & 18 & 22416   \\
      &            & 2010-04-20, 2010-04-22, 2010-05-16, 2010-05-18,   & &  \\
      &            & 2010-06-14, 2010-07-10, 2010-08-05, 2010-09-02,  & & \\
      &            & 2010-09-26, 2010-09-30, 2010-10-25, 2010-11-17,   & &\\
      &            & 2010-11-21, 2010-12-16   &&\\
\hline
2     &  XRT       & 2006-09-05, 2006-11-21 - 2006-11-24, 2006-12-18, & 10 & 20147 \\
Superorb. phase : 0.5 -- 0.7       &            & 2006-12-20, 2006-12-22, 2011-01-14, 2011-10-01 &&\\
Orbital phase : 0.6 -- 0.8      &  PCA       & 2006-10-27, 2006-10-29, 2011-01-09, 2011-01-13, & 20 & 24384 \\
      &            & 2011-02-06, 2011-02-09, 2011-03-06, 2011-03-31, && \\
      &            & 2011-04-03, 2011-04-28, 2011-05-22, 2011-05-26, && \\
      &            & 2011-06-19, 2011-07-13, 2011-07-17, 2011-08-10, && \\
      &            & 2011-08-14, 2011-09-07, 2011-10-03, 20011-10-30 && \\
\hline
3     &   XRT      &  2008-10-22, 2008-11-19, 2008-12-17, 2013-11-23, & 6 & 10266 \\
Superorb. phase : 0.0 -- 0.2       &            &  2013-12-14, 2014-01-11 && \\
Orbital phase : 0.0 -- 0.2       &   PCA      &  2008-10-22, 2008-10-25, 2008-11-17, 2008-11-19, & 21 & 32912 \\
      &            &  2008-12-13, 2008-12-17, 2009-01-10, 2009-02-04,   && \\
      &            &  2009-02-07, 2009-03-05, 2009-03-29, 2009-04-02,   && \\
      &            &  2009-04-26, 2009-05-21, 2009-05-25, 2009-06-18,   && \\
      &            &  2009-07-12, 2009-07-15, 2009-08-08 &&\\
\enddata
%\label{tab:xrt-pca-log}
\end{deluxetable*}

%\clearpage
A combined spectral fit was performed for XRT and PCA data. The PCA spectrum was
normalized with the XRT spectrum for this purpose. The XRT and PCA spectra covering the energy range
of 0.7-20 keV were fitted by using XSPEC with a powerlaw with the line-of-sight absorption,
which was kept free during the fit.
Model
parameters for the combined fit as well as for only XRT data are listed in Table
\ref{tab:fit-params-xrt-pca}.
Since the bandwidth of the data is quite limited, we find a correlation
between the power-law index and the absorption, indicating that a steeper power-law
is compensated by a large absorption. For the wide band fitting, we use the joint
XRT-PCA fit because the higher energy data from PCA    constrains the power-law better.

\begin{deluxetable}{cccc}
%\tabletypesize{\scriptsize}
\tablecaption{Best-fit parameters of a power-law (with absorption) fit to the data for XRT and PCA \label{tab:fit-params-xrt-pca}}
\tablewidth{0pt}
\tablehead{
\colhead{}  &\multicolumn{3}{c}{Only XRT}\\
\colhead{}  & \colhead{$N_H$ (10$^{22}$ cm$^{-2}$)}  & \colhead{alpha}  &  \colhead{norm}\\
}
\startdata
state1    & 0.68$\pm$0.05  & 1.58$\pm$0.06 &   (2.51$\pm$0.21)$\times 10^{-3}$  \\
state2    & 0.70$\pm$0.05  & 1.53$\pm$0.05  &  (3.08$\pm$0.20)$\times 10^{-3}$   \\
state3    & 0.69$\pm$0.11  & 1.47$\pm$0.12  &  (1.25$\pm$0.20)$\times 10^{-3}$    \\
\hline
& \multicolumn{3}{c}{XRT+PCA (all layers)} \\
&  $N_H$ (10$^{22}$ cm$^{-2}$)      &     alpha      &      norm     \\
\hline
state1   & 0.81$\pm$0.04  & 1.79$\pm$0.03  &  (3.22$\pm$0.15)$\times 10^{-3}$ \\
state2   & 0.90$\pm$0.03  & 1.78$\pm$0.03  &  (4.27$\pm$0.17)$\times 10^{-3}$  \\
state3   & 1.05$\pm$0.08  & 1.95$\pm$0.05  &  (2.25$\pm$0.18)$\times 10^{-3}$  \\
\enddata
\end{deluxetable}

Fermi-LAT SEDs for the three states fitted with a cutoff powerlaw are given in Fig.
\ref{fig:fermi-sed} and model parameters are listed in Table \ref{tab:fermi-param}.
Some differences are seen in the spectral indices for \gam-rays between state1
and other states (see Table \ref{tab:fermi-param}). In case of X-ray data some 
steepening of the spectrum is seen as flux decreases, as indicated by variation in
spectral index  (see Table \ref{tab:fit-params-xrt-pca}).

\bfig[h]
\centering
\includegraphics[scale=0.45]{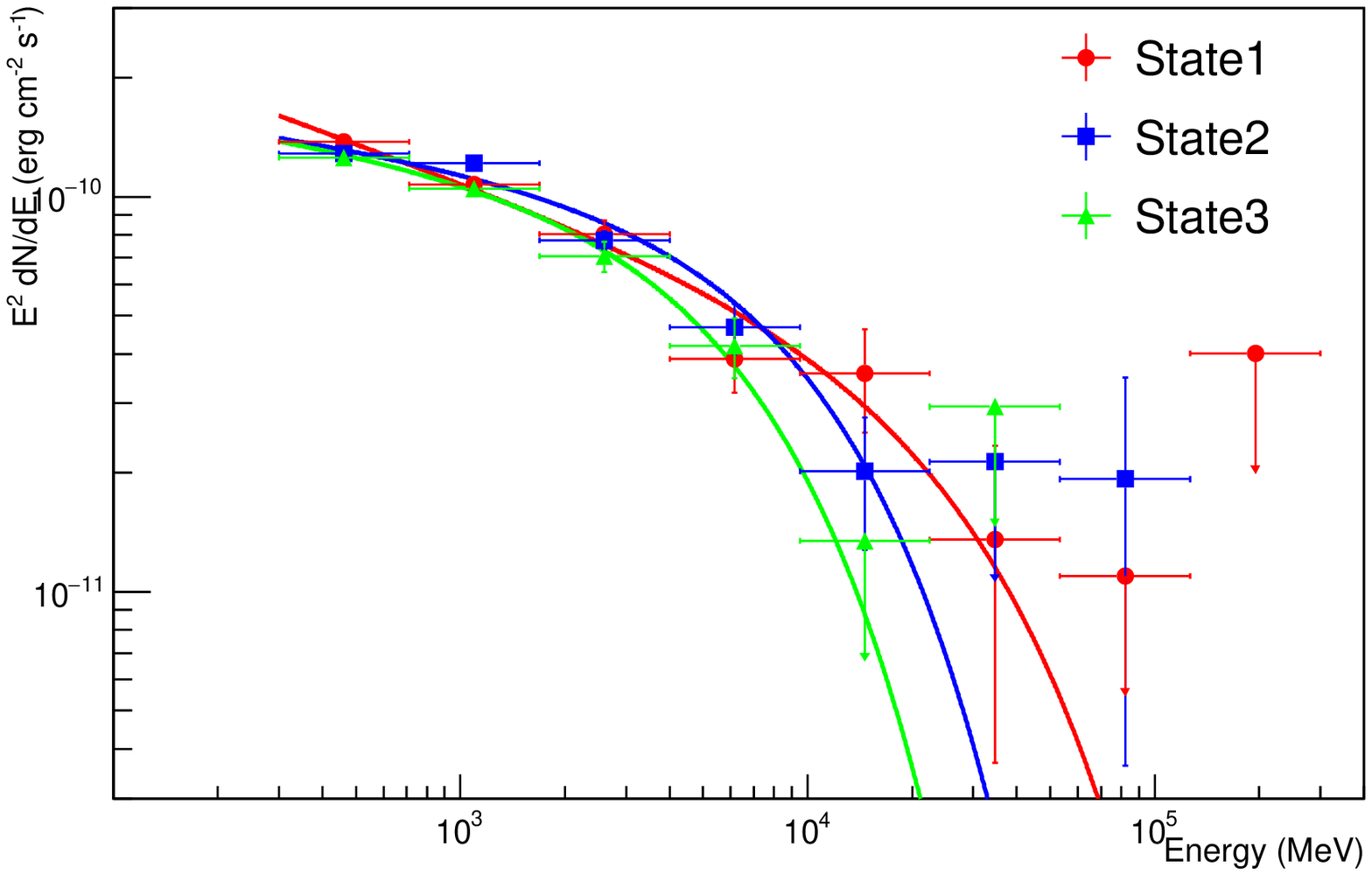}
\caption{A cutoff power law fit to Fermi-LAT data for the three different states. Best-fit curves are shown
as solid lines.}
\label{fig:fermi-sed}
\efig

\begin{deluxetable}{cccc}
\tablecaption{Parameters of a cutoff power law fit to the Fermi-LAT data for  three different states \label{tab:fermi-param}}
\tablewidth{0pt}
\tablehead{
\colhead{parameters} & \colhead{state1} & \colhead{state2} & \colhead{state3}
}
\startdata
$\alpha$                        &   2.31         &   2.12       &  2.12    \\
flux                            &   2.48 $\times 10^{-7}$       &   2.41$\times 10^{-7}$    & 2.28$\times 10^{-7}$  \\
Ec (MeV)                        &   30041         &  10000          &  6338   \\
TS                              &   2663         &  2727          &  2328    \\
\enddata
\end{deluxetable}

We have investigated the   spectral energy distributions (SEDs) of the source. The state3 
does not have TeV data and hence we have made a detailed SED study for the
other two states. These states are bright in all wavebands and hence can be used as
a template to understand the emission mechanisms. 
VERITAS spectral data obtained from \cite{Acciari-2011} corresponds to state1.
For radio flux, the average of 15 GHz data from OVRO described in section 2 
is used. This sets an upper limit on the modelled radio flux. In addition, 
we have also plotted radio data from \citet{Strickman_1998}, which correspond to orbital 
phase of 0.8 and superorbital phase of 0.8.

Since \lsi\ is identified as a potential microquasar based on radio observations, 
high energy emission is likely to be produced in jets.  In case of microquasars, 
compact object could be a neutron star or a black hole accreting matter from a
companion star which presumably drives relativistic outflow or jet from the
compact object. Acceleration of charged particles in the jet produces high
energy emission. We have considered this scenario to model the SEDs. In the
context of leptonic model, the low energy emission arises from Synchrotron emission
from ultra-relativistic electrons in the jet. Whereas the high energy emission
arises from inverse Compton scattering of soft photons, which could be either
soft photons from Synchrotron radiation (Synchrotron Self-Compton or SSC 
model) or photons from companion star or accretion disk (External Compton
model). In this work, relativistic jet making an angle of 30$^\circ$ 
(\cite{Gupta-2006} and reference therein) with our line of sight is considered.
Electrons are assumed to have a broken power-law energy spectrum given by

\begin{eqnarray}
{dn_e \over d\gamma}  \propto \begin{cases}
 \gamma^{-\alpha} ~\mbox{for} ~\gamma <\gamma_{br} \nonumber \\
  \gamma^{-\beta} \exp{\left(-{\gamma \over \gamma_c} \right )} ~ \mbox{for}~ \gamma_{br}\leq \gamma \leq \gamma_c                   
  \end{cases}
\label{eqn:pi0}
\end{eqnarray}

where, $n_e$ denotes the number density of the electrons, $\gamma$ is 
the Lorentz factor of the electron, $\alpha$ and $\beta$ are spectral indices,
$\gamma_{br}$ break energy and $\gamma_c$ the highest energy of the electron.

\begin{figure}
\centering
\includegraphics[scale=0.45]{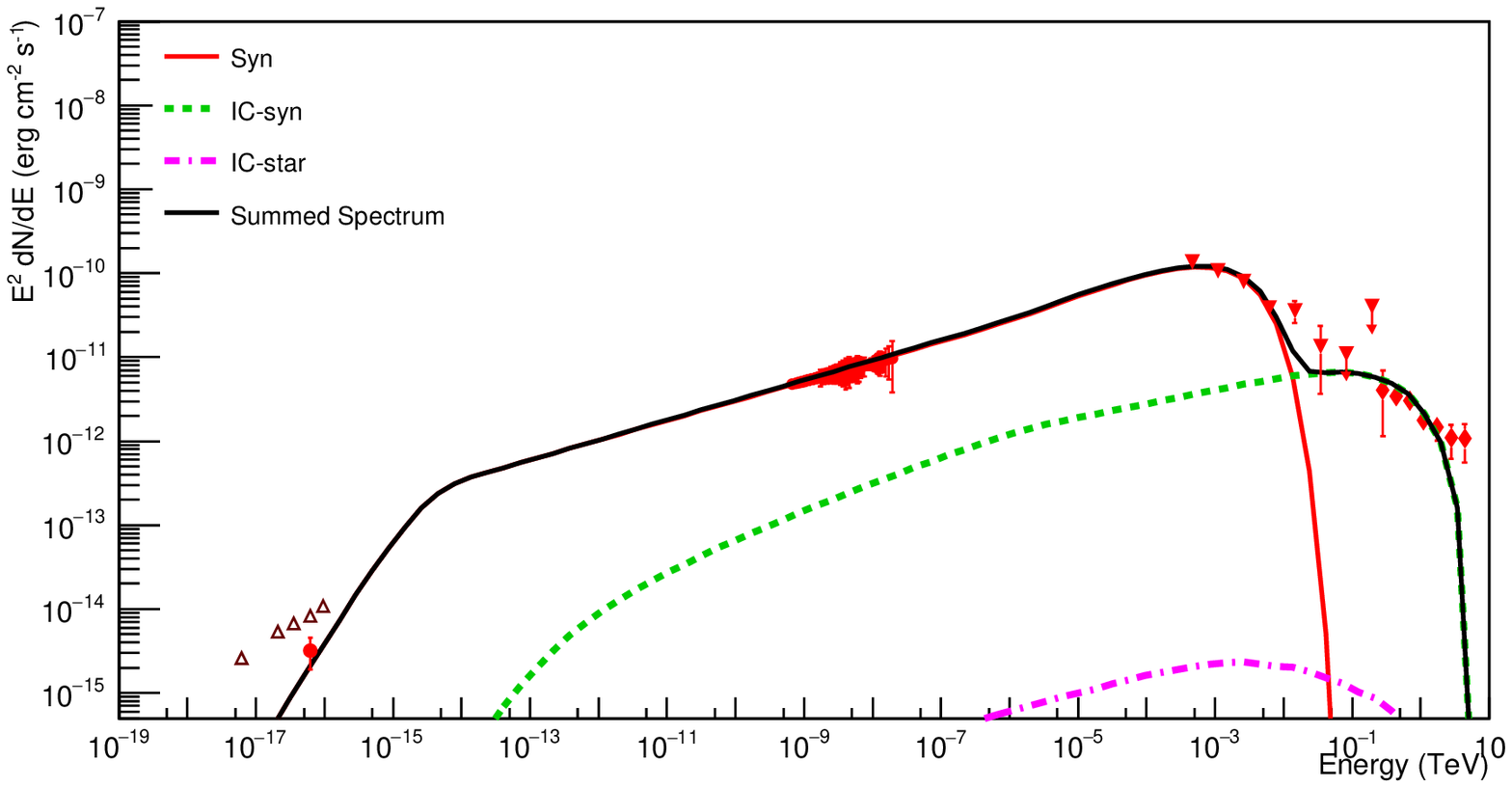}
\caption{The SED of \lsi\ for state1.
The synchrotron and inverse Compton spectra are calculated using the  parameters
as given in Table \ref{tab:params_fit_SED}. X-ray, Fermi-LAT and 
VERITAS data for state 1 are shown with points in red color. Radio data shown in
the figure do not correspond to state 1. The average flux from OVRO is shown with 
filled circle of red color, whereas radio data from \citet{Strickman_1998} is shown 
with brown triangles.}
\label{fig:xxx_p}
\end{figure}

\begin{figure}
\centering
\includegraphics[scale=0.45]{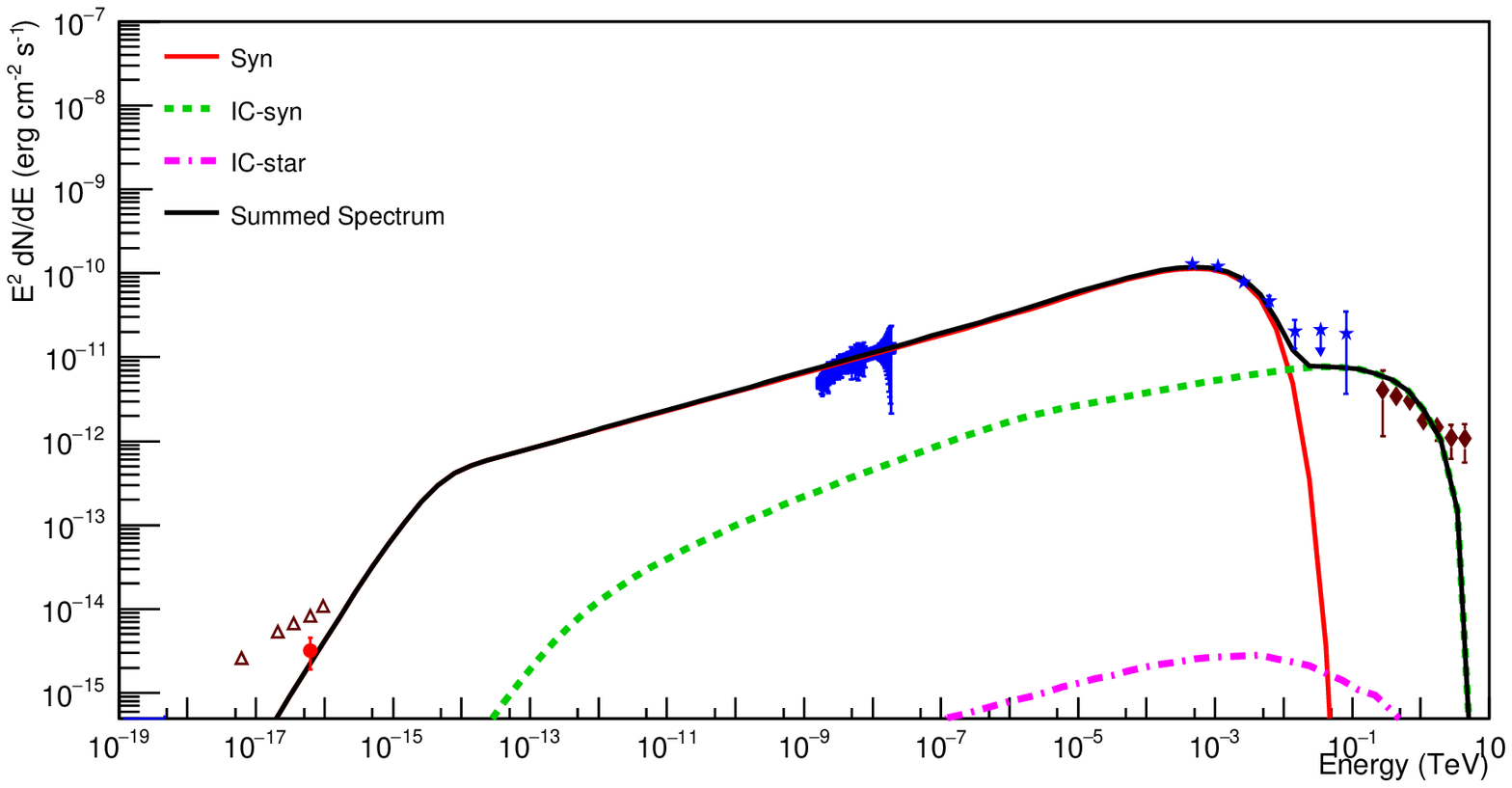}
\caption{The SED of \lsi\ for state2.
The synchrotron and inverse Compton spectra are calculated using the same parameters
as given in Table \ref{tab:params_fit_SED}. X-ray and Fermi-LAT data 
for state 2 are shown with points in blue color. Since, VERITAS data for state 2 is 
not available, state 1 VERITAS data is used which is shown in brown colour. Radio 
data are the same as in Fig.~\ref{fig:xxx_p}.}
\label{fig:xxx_a}
\end{figure}

For this source, the distance is taken as 2 kpc \citep{Hutchings-1981,
Frail-1991} and Lorentz factor for bulk motion is assumed to be
1.25 \citep{Massi-2004}. The models are shown for state1 and state2 respectively
in Figure \ref{fig:xxx_p} and Figure \ref{fig:xxx_a}. 

\begin{figure}
\centering
\includegraphics[scale=0.45]{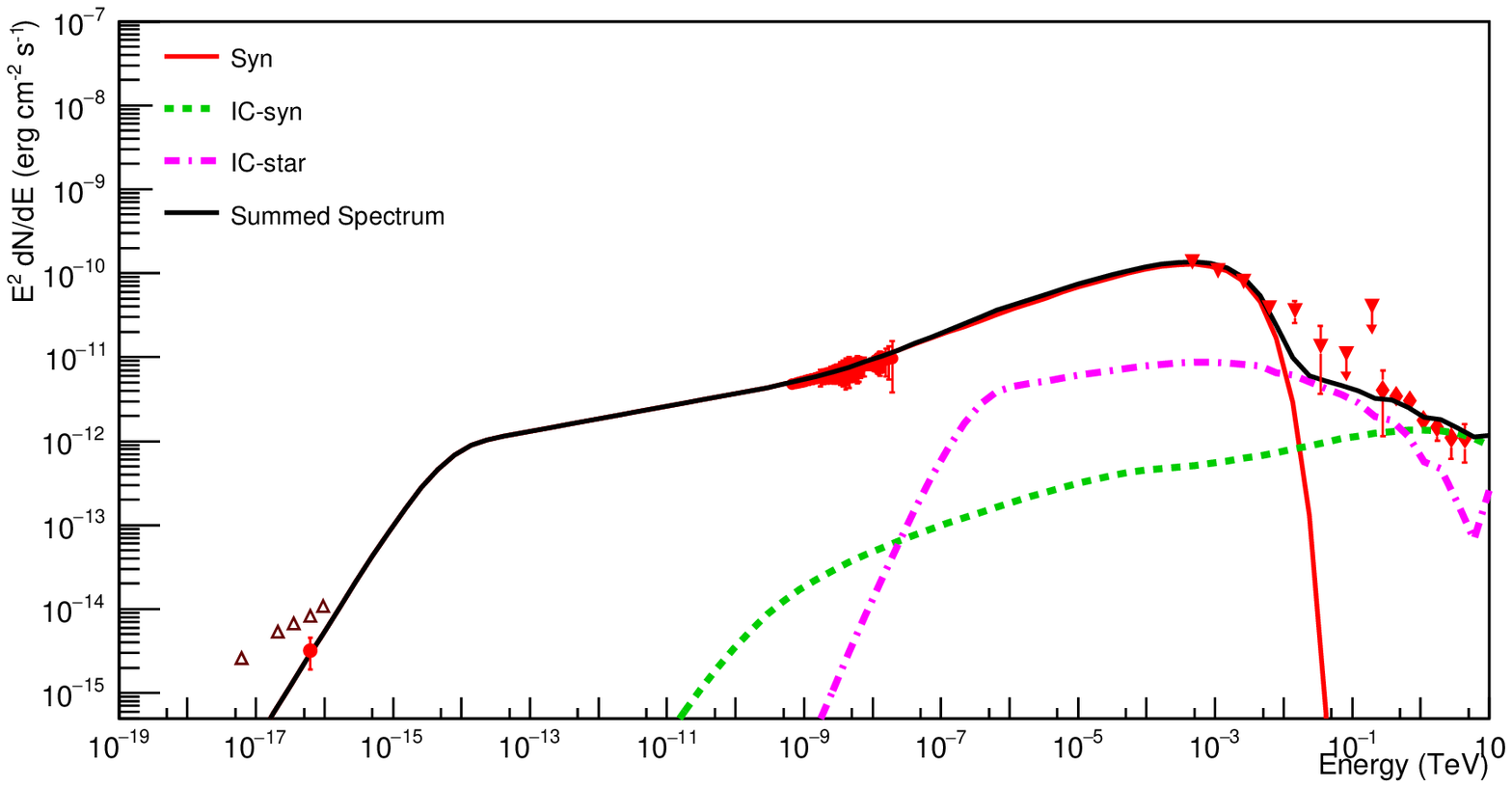}
\caption{The SED of \lsi\ for state1.
The synchrotron and inverse Compton spectra are calculated using the same parameters
given in the last column of Table \ref{tab:params_fit_SED}, with the emission region radius
decided from the variability time scale.}
\label{fig:xxx_new_B.eps}
\end{figure}

Here it is assumed that the radio, X-ray and \gam-ray emissions originate in the
same region and the magnetic field in emission blob is quite high, of the
order of $10^3$ G. Rest of the model parameters are fitted 
and these parameters are listed in Table \ref{tab:params_fit_SED}.
To explain the TeV \gam-ray emission it was necessary to include IC of
photons from accretion disk or companion star in addition to the SSC
component. Radiation density ($U_{rad}$) is estimated from luminosity  
$L$ using expression $U_{rad} = L/  {4 \pi R^2 c}$, where $R$ is the 
distance of the emission volume from the companion star or the accretion disk. 
Radiation density 
from the companion star, with $L_c = 2 \times 10^{38}$ erg s$^{-1}$ and a distance of 
$R~\sim 10^{12}$ cm, is about  4 orders of magnitude higher than the
corresponding radiation density from the  accretion disk. Hence, we have considered 
only the seed photons from the companion star for the External Compton model. However, 
this spectrum cannot explain the observed data as seen from Figure \ref{fig:xxx_p} 
and Figure \ref{fig:xxx_a}. In this case, we have considered radius of the emission 
volume as a parameter for the fit to the data. We can also estimate the radius of 
emission volume from the variability time scale of the source. We fixed the size 
of the emission region according to the estimates from variability study \citep{Smith-2009},
which indicates a possible size of the emission region to be $\sim 6 \times 10^{10}$ cm.
Considering that the bulk Lorentz factor is 1.25, this size   corresponds to
$\sim~7.5 \times 10^{10}$ cm. Fixing the emission region size to this value, 
model parameters were estimated which are given in the last column of Table \ref{tab:params_fit_SED}.
Although the synchrotron spectrum explains the observed fluxes from radio to MeV--GeV 
energies, SSC spectrum alone cannot fit the data well. Hence we have also estimated 
the contribution of companion star photons for this low magnetic field case and we 
found that the external Compton model overestimates the observed flux at MeV--TeV 
energies. However, SSC and EC models together can explain the data well if the 
companion star luminosity is considered to be reduced by a factor of 10. This is 
shown in Figure \ref{fig:xxx_new_B.eps}.

In the spectral fitting, we did not consider the radio data (triangles 
in Fig.~\ref{fig:xxx_a}, \ref{fig:xxx_p} and \ref{fig:xxx_new_B.eps}) from VLA observation 
\citep{Strickman_1998} in fit, since the orbital and superorbital phases for these are 
different from the phases for state 1 and state 2. Since, the energy spectrum is not 
available for OVRO data, we have used average flux as an upper-limit for SED modelling, 
and for the chosen set of parameters the model does not overestimate radio fluxes for 
the states considered above.

\begin{deluxetable}{cccc}
\tablecaption{Parameters of the fit for microquasar scenario \label{tab:params_fit_SED}}
\tablewidth{0pt}
\tablehead{
\colhead{Parameters} & \colhead{state1} & \colhead{state2} & \colhead{state1 (radius from} \\
\colhead{} & \colhead{} & \colhead{} & \colhead{variability study)} \\
}
\startdata
 Magnetic field (Gauss)  & 5 $\times 10^3$   & 5 $\times 10^{3}$  & 15 \\
$\gamma_{min}$          & 4.4               & 4.9              & 110 \\
 $\gamma_{max}$          & 5.6 $\times 10^6$        & 5.4 $\times 10^6$  &  9.0 $\times 10^7$ \\
 %Electron density         & 0.8e57            & 1.0e51               \\
 spectral index ($\alpha$) & 2.53             & 2.55          & 2.7       \\
 spectral index ($\beta$ )  & 2.34            & 2.40           & 2.4      \\
 radius (cm)                 & 11.5 $\times 10^7$ &   18.0 $\times 10^7$   & 7.5 $\times 10^{10}$             \\
 Gamma Break            & 1.0 $\times 10^5$        & 1.4 $\times 10^7$   & 9.0 $\times10^4$       \\
 Bulk Lorentz factor    & 1.25              &  1.25         & 1.25        \\
 Distance (kpc)         & 2.0               & 2.0              & 2.0   \\
Inclination angle(deg)  & 30.0              & 30.0       & 30.0          \\
 %density (n)               & 15                & ---                \\
 Luminosity (erg/s)     & 3.9 $\times 10^{35}$ &  3.8 $\times 10^{35}$ &  4.3 $\times 10^{35}$ \\
\enddata
\end{deluxetable}

\section{Discussion and Conclusions}\label{sec:discussion}

Long-term timing analyses of \lsi\ at different wavelengths has shown some of the 
interesting characteristics of the source. Flux in various wavebands shows 
variation with superorbital phase and this variation is wavelength dependent as well
as the binary phase dependent.
At X-ray energies, as evident from Figure \ref{fig:density-plot}, the source is
bright at orbital phases $\sim 0.4-0.8$ and superorbital phases of $\sim
0.3-0.8$. Whereas at radio energies, the source is bright at orbital phases of
$\sim 0.4-0.8$ and superorbital phases of $\sim 0.7-1.4$.    
The \gam-ray flux in MeV-GeV band as given by Fermi-LAT
shows a shift relative to the radio and the X-ray bands.
This behaviour possibly indicates that radio,
X-ray and \gam-ray emissions could be originating from different regions.

The long-term superorbital modulation of flux could support the scenario where 
circumstellar disk of a Be star quasi-cyclically expands and shrinks (e.g., 
\cite{Negueruela-2001}). However, for such a scenario the long-term period is 
variable from cycle to cycle \citep{Rivinius_2013}. Recent analysis of radio 
data established the fact that the long-term period is quite stable over 8 
cycles \citep{Massi_2016} which makes the scenario of quasi-cyclic variation 
of circumstellar disk of the Be star for \lsi\ less probable. This stable 
superobital modulation is attributed to periodic Doppler boosting effects 
of the precessional jets associated with the compact objects \citep{Massi_2014}.

In this paper, we have seen that the modulation of flux with superorbital phase 
is more prominent in orbital phase bins near apastron. This is clearly seen at 
various wavelengths in Figures \ref{fig:xrt-lc2} and \ref{fig:super_peak_orb}. 
Although the long-term superorbital variation does not support the variation of 
circumstellar disk size,  this type of superorbital modulation near the apastron 
could stem from the interaction of the compact object with the circumstellar disk 
of the Be star. The equivalent width (EW) of the H$\alpha$ emission line is 
related to the size of the stellar disk \citep{Zamanov-Mart-2000,Grundstrom_2006}. 
In addition to that, it has been found that the maximum of the EW of H$\alpha$ 
occurs in a region around superorbital phase of $\sim$ 0.4 (see \cite{Zamanov-1999,
Zamanov-Mart-2000}) considering superorbital period of 1584 days. However, if we 
use superorbital period as 1626 days then  the maximum of the EW of H$\alpha$ 
occurs  at $\sim$ 0.3. From Figure  \ref{fig:super_peak_orb} we see that flux 
of gamma-rays is high at the superorbital phase of $\sim$ 0.3--0.5, which suggests 
that the disk plays an important role
in modulating $\gamma$-rays. Although, a similar enhancement of X-rays at superorbital
phase of 0.2 is seen by \cite{Li-2011} considering only peak flux, we see that X-ray
flux peaks at the superorbital phases in the range  of $\sim$ 0.4--0.8 depending on the
orbital phase. We see that the peak of radio flux is shifted further. It suggests that even 
if the disk size plays a significant role for $\gamma$-rays, X-ray and radio fluxes 
are not necessarily  affected much by the size of the disk.

Figure \ref{fig:xrt-lc2} shows that, for all wavebands, the superorbital variability 
is not significant in the periastron region, whereas it is significant at the apastron.
This can support the scenario where one assumes that the interaction between
compact object and the circumstellar disk of Be star is strong when compact object is in the
proximity of Be star.
As a result, superorbital modulation effect becomes insignificant as 
suggested by \cite{Ackermann-2013}.  However, it becomes dominant as the compact object 
starts moving towards the apastron region. 

Another possible scenario for the modulation is related to the precession of the Be star disk about 
the orbital  plane. If this scenario is adopted for possible 
explanation of the 
strong superorbital modulation in the apastron phase, then the angular distance between 
orbital plane and the disk plane should become minimum. As a result, even if the compact 
object is far from the Be star, the smaller angular distance between disk plane and orbital 
plane provides relatively higher interaction of the compact object with the disk.   

In addition to the superorbital modulation in the apastron phase (0.5 -- 1.0) we have seen 
phase lag among radio, X-ray and $\gamma$-rays.  The possible explanation for the constant 
phase lag between X-ray and radio is  that the plasma blobs filled with 
high-energy particles may escape from the X-ray emission region to the radio emission region 
which is at a distance of $\sim$  
10 times the binary separation distance as proposed by \cite{Chernyakova-2012} in the context 
of pulsar wind scenario. However, in the microquasar scenario, different regions in the jets can be 
responsible for the phase lag.  We have also seen the phase lag between radio and 
$\gamma$-rays. In such binary systems, $\gamma$-rays are considered to be produced by 
up-scattering of radio photons or accretion disk/star photons. If the $\gamma$-rays are 
originating through the up-scattering of radio photons which are being produced by the same 
population of electrons then there  should not be any phase lag between radio and 
$\gamma$-rays. Hence, up-scattering of  a separate population of photons could be a
possible explanation for the phase lag between $\gamma$-ray and radio fluxes. 

In addition to the timing analysis, we also tried to understand the spectral behaviour of the 
source at different orbital and superorbital phases. We have chosen three 
different regions following flux variations for X-ray, radio, and \gam-rays as function of 
orbital and superorbital phase. From Figure \ref{fig:density-plot} we see that the source 
at high energy is mostly very 
active in the orbital phase bin of 0.5--0.8 and superorbital phase bin 0.3--0.7.  We 
selected two different regions  with superorbital phase 0.3--0.5; orbital phase: 0.6--0.8 (state1) 
and  superorbital phase: 0.5 -- 0.7; orbital phase: 0.6 -- 0.8 (state2) from this region where source 
is bright at all wavelengths. To compare the spectral variation with the other orbital and 
superorbital phases where the source is not bright, we have chosen a region with  superorbital 
phase 0.0 -- 0.2 and orbital phase 0.0 -- 0.2 (state3). 
Based on these three different regions of orbital or superorbital phases, we have analysed 
X-ray and Fermi-LAT data to see the spectral behaviour of the source at high energies. 
We found no significant differences  between flux levels but we see some variations in the
spectral indices at Fermi-LAT energies. However, we see some difference in both the spectral
indices and flux levels for XRT-PCA data, though the interplay of the spectral shape and the absorption
 playing a role in this trend cannot be ruled out.

From the fit to the SED we have seen that we can explain the data well considering \lsi\ as a microquasar. 
In the microquasar model, it is generally assumed that the high energy emission comes from the region which 
is very close to the compact object to reduce the effect of $\gamma \gamma$ absorption due to 
radiation field of companion star \citep{Gupta-2006_a}. Magnetic field in this region is relatively high 
as considered 
for our model, and we have estimated emission volume of the order of $~10^{8}$ cm. In this emission 
volume, some the of the emitted $\gamma$-rays can be absorbed  through $e^+e^-$ pair creation process due to X-ray 
photons in  emission volume. We have estimated that about 20\% of $\gamma$-rays will be absorbed at TeV energies. 
However, larger sizes of emission volume will make this absorption insignificant. 
We have seen that it is possible to have lower values of magnetic field strength 
to explain the observed data, in case of larger emission volume. 

We have also seen in section \ref{sec:sed} that if we consider the radius of emission volume obtained
from variability study, the magnetic field from the model fit to the data is estimated to be $\sim 10$ G. 
However, we found that the SSC model alone cannot explain the TeV data well and EC model overestimates 
the observed flux for the luminosity of the companion star $\sim 10^{38}$ erg s$^{-1}$. A lower value 
of this luminosity ($\sim 10^{37}$ erg s$^{-1}$) can explain the data well. This suggests that lower values of 
magnetic field in the emission blobs are suitable for \lsi\ to explain the observed data constraining 
the luminosity of the companion star. With high magnetic field, the Synchrotron cooling timescale is 
much smaller than the variability timescale which could be as low as 2 seconds as estimated by 
\cite{Smith-2009}. In our SED fitting, we have considered that the emitting blob is close to the compact 
object.  The blob size increases as it moves away from the compact object in the jets and magnetic 
field reduces. The time-averaged values of flux from a particular region in the  jet as considered 
by \cite{Gupta-2006_a} could reduce the discrepancy between Synchrotron cooling time scale and the time 
scale of X-ray flux variability. From the SED, it 
seems that the one single emission process is responsible for X-ray and MeV-GeV data. Hence, we have 
considered Synchrotron emission process to explain the data up to GeV energies. As a result, a high 
magnetic field is required to explain the data if the maximum energy of high energy electrons is not 
well above $\sim$ 1 TeV. A good quality data in the hard X-ray region can establish whether a different 
emission component is  required to explain the data at MeV-GeV region. It can also indicate if we need 
a different population of electrons to explain data at different energy bands in the SED.

We have also seen that the fitted model parameters show hardening of spectral index after break (see 
Table \ref{tab:params_fit_SED}). In addition, flux levels for different states (mainly X-ray) are 
different. A change in the location of the compact object relative to the companion star during the orbital and the 
superorbital cycles and its interaction with circumstellar disk could be responsible for the changing 
electron spectral distribution.

In the context of timing analysis, we have seen the phase lag among radio, X-ray and \gam-ray data which 
may suggest that they originate from different emission regions. However, in our present spectral modelling we have 
considered single emission zone to explain the mutiwavelength data. To support the scenario of different origins we 
need simultaneous multiwavelength data for longer period both for timing and spectral analysis. At present, 
we have such observation for radio, X-ray and MeV--GeV gamma-rays. However, GeV--TeV data is also required to get 
a complete understanding of the source in multifrequencies.

The following major conclusions can be drawn, based on the study presented here:

\begin{itemize}
\item
{The super orbital modulation is more pronounced near the apastron for all 
wavelengths, supporting geometric scenarios as a cause for the super orbital modulation.
}
\item
{There is a definite wavelength dependent variation of the maximum of the super orbital
flux with respect to the binary phase. This variation shows a wavelength dependent
shift.
}
\item
{Emission from radio to GeV gamma-rays during the maximum emission can be modelled by an one-zone micro-quasar 
jet model.  To explain the TeV emission, Comptonization from
an External Compton source is necessary especially when low magnetic field is assumed. 
In this case, we suggest that the photons from the companion star, with a lower luminosity
($\sim$10$^{37}$ erg s$^{-1}$), is adequate to explain the data.
}
\item
{Extended hard X-ray data would be necessary to constrain the synchrotron model and 
TeV observations across a super orbital cycle, along with X-ray measurements, would
be required to make a detailed emission model for this source.
}
\end{itemize}

\section*{Acknowledgements}

We acknowledge the use of data from the High Energy Astrophysics Science Archive
Research Center (HEASARC), provided by NASA's Goddard Space Flight Center. Also the data
supplied by the UK Swift Science Data Centre at the University of Leicester has been used
in present work.
We thank Hovatta Talvikki for providing us data from OVRO 40-m monitoring program which was used in 
the research by \citep{Richards-2011}(supported in part by NASA grants NNX08AW31G 
and NNX11A043G, and NSF grants AST-0808050 and AST-1109911). We acknowledge the use of Fermi-LAT data 
and analysis tool from Fermi Science Support Center. We would also like to thank MAGIC collaboration 
for making their published data public which has been used in this work. We also acknowledge VERITAS
collaboration for their published data used in this work.

\bibliographystyle{apj}
\bibliography{LSI_apj_manuscript_R2_v2.bib}

\end{document}